\documentclass[sigconf]{acmart}
\makeatletter
\def\@ACM@checkaffil{
    \if@ACM@instpresent\else
    \ClassWarningNoLine{\@classname}{No institution present for an affiliation}%
    \fi
    \if@ACM@citypresent\else
    \ClassWarningNoLine{\@classname}{No city present for an affiliation}%
    \fi
    \if@ACM@countrypresent\else
        \ClassWarningNoLine{\@classname}{No country present for an affiliation}%
    \fi
}
\makeatother
\usepackage{graphicx}
\usepackage{textcomp}
\usepackage{xcolor}
\usepackage{booktabs} 
\usepackage{amsmath}
\usepackage{amsfonts}
\usepackage{comment}
\usepackage{footmisc}
\usepackage{mathrsfs}
\usepackage{graphicx}
\usepackage{subfigure}
\usepackage{multirow}
\usepackage{algorithm}
\usepackage{algorithmic}
\usepackage{multicol}  
\usepackage{enumitem}
\usepackage{array}
\usepackage{float}
\usepackage{hyperref}
\usepackage{xspace}
\hypersetup{hidelinks}
\newcommand{\etal}{\emph{et al.}\xspace}

\newcommand{\ie}{\emph{i.e.,}\xspace}

\AtBeginDocument{%
  \providecommand\BibTeX{{%
    \normalfont B\kern-0.5em{\scshape i\kern-0.25em b}\kern-0.8em\TeX}}}

\setcopyright{acmcopyright}
\copyrightyear{2023}
\acmYear{2023}
\setcopyright{acmlicensed}\acmConference[WWW '23]{Proceedings of the ACM Web Conference 2023}{May 1--5, 2023}{Austin, TX, USA}
\acmBooktitle{Proceedings of the ACM Web Conference 2023 (WWW '23), May 1--5, 2023, Austin, TX, USA}
\acmPrice{15.00}
\acmDOI{10.1145/3543507.3583440}
\acmISBN{978-1-4503-9416-1/23/04}

\begin{document}

\title{AutoMLP: Automated MLP for Sequential Recommendations}

\author{Muyang Li}
\affiliation{%
  \institution{City University of Hong Kong}
  \institution{University of Sydney}
}
\email{muli0371@uni.sydney.edu.au}
\authornote{Both authors contributed equally to this research.}
\author{Zijian Zhang}
\affiliation{%
  \institution{City University of Hong Kong}
  \institution{Jilin University}
}
\email{zhangzj2114@mails.jlu.edu.cn}
\authornotemark[1]
\author{Xiangyu Zhao}
\affiliation{%
  \institution{City University of Hong Kong}
}
\email{xianzhao@cityu.edu.hk}
\authornote{Xiangyu Zhao is the corresponding author.}
\author{Wanyu Wang}
\affiliation{%
  \institution{City University of Hong Kong}
  \country{}
 }
 \email{wanyuwang4-c@my.cityu.edu.hk}
\author{Minhao Zhao}
\affiliation{%
  \institution{Fuxi AI Lab, NetEase}
}
\email{zhaominghao@corp.netease.com}
\author{Runze Wu}
\affiliation{%
  \institution{Fuxi AI Lab, NetEase}
}
\email{wurunze1@corp.netease.com}
\author{Ruocheng Guo}
\affiliation{%
  \institution{Bytedance AI Lab}
}
\email{rguo.asu@gmail.com}


\begin{abstract}
Sequential recommender systems aim to predict users' next interested item given their historical interactions. However, a long-standing issue is how to distinguish between users' long/short-term interests, which may be heterogeneous and contribute differently to the next recommendation. Existing approaches usually set pre-defined short-term interest length by exhaustive search or empirical experience, which is either highly inefficient or yields subpar results. The recent advanced transformer-based models can achieve state-of-the-art performances despite the aforementioned issue, but they have a quadratic computational complexity to the length of the input sequence. To this end, this paper proposes a novel sequential recommender system, AutoMLP, aiming for better modeling users' long/short-term interests from their historical interactions. In addition, we design an automated and adaptive search algorithm for preferable short-term interest length via end-to-end optimization. Through extensive experiments, we show that AutoMLP has competitive performance against state-of-the-art methods, while maintaining linear computational complexity. 
\end{abstract}



\maketitle

\section{Introduction}
\label{sec:introduction}

\begin{figure}
	\centering
	\includegraphics[width=1\linewidth]{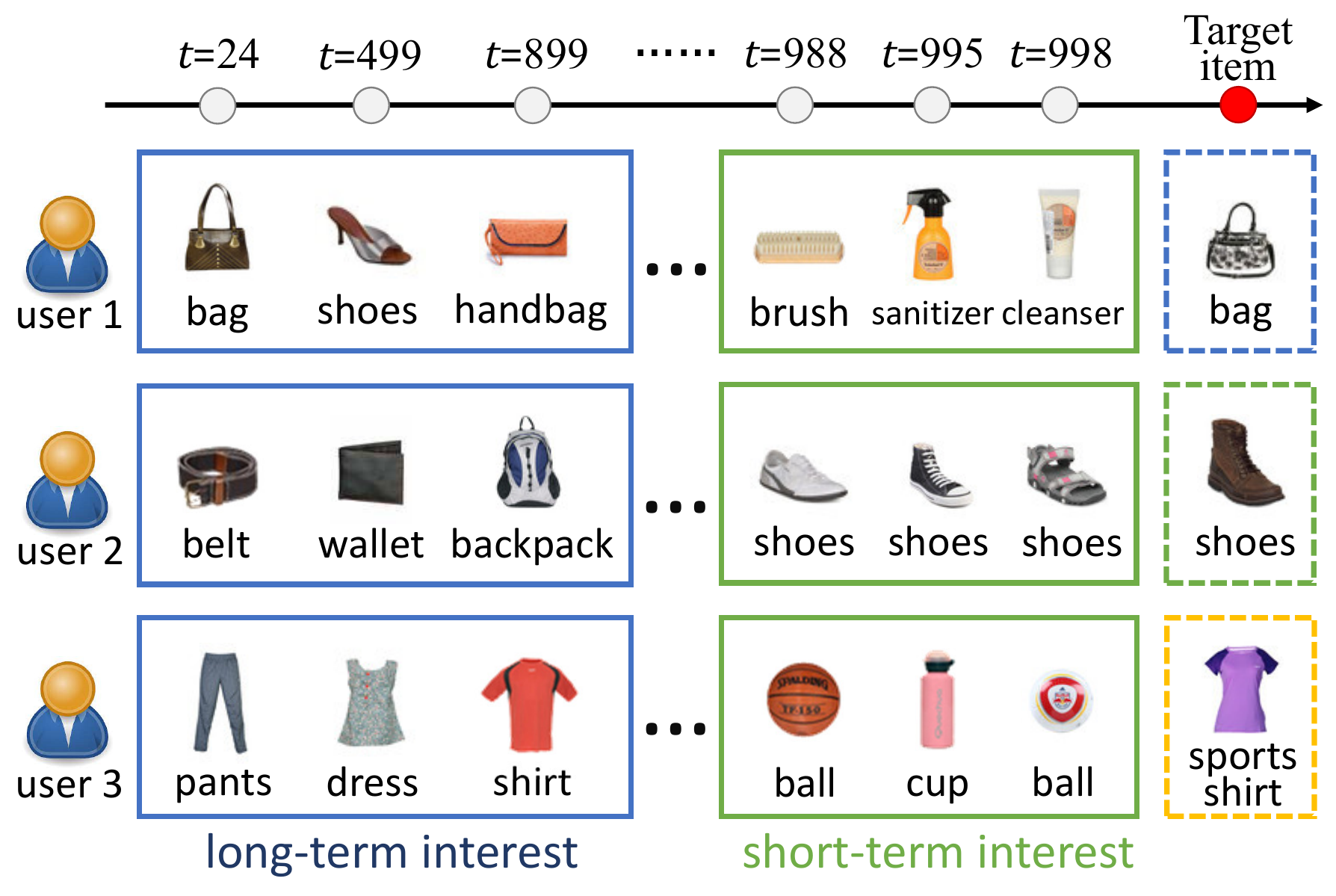}
	\caption{Illustrative examples of users' long/short-term interests. The target items of user 1 and 2 are related to their long- and short-term interests respectively, and user 3 is influenced by both. }
	\label{fig:Example}
    \vspace*{-5mm}
\end{figure}

Sequential recommender systems have been playing an essential role in real-world service providers, such as purchase prediction \cite{zhang2019feature, zhou2019deep, yu2019multi}, web page recommendations \cite{hu2020graph}, and next point-of-interest recommendations \cite{sun2020go, feng2018deepmove, islam2022survey}. 
It models the sequential dependency based on users' historical behaviors, which consistently outperforms static recommendation models due to the utilization of sequential information \cite{quadrana2018sequence, hidasi2015session, hidasi2018recurrent}.

\begin{figure*}[t]
	\centering
	\includegraphics[scale=0.966]{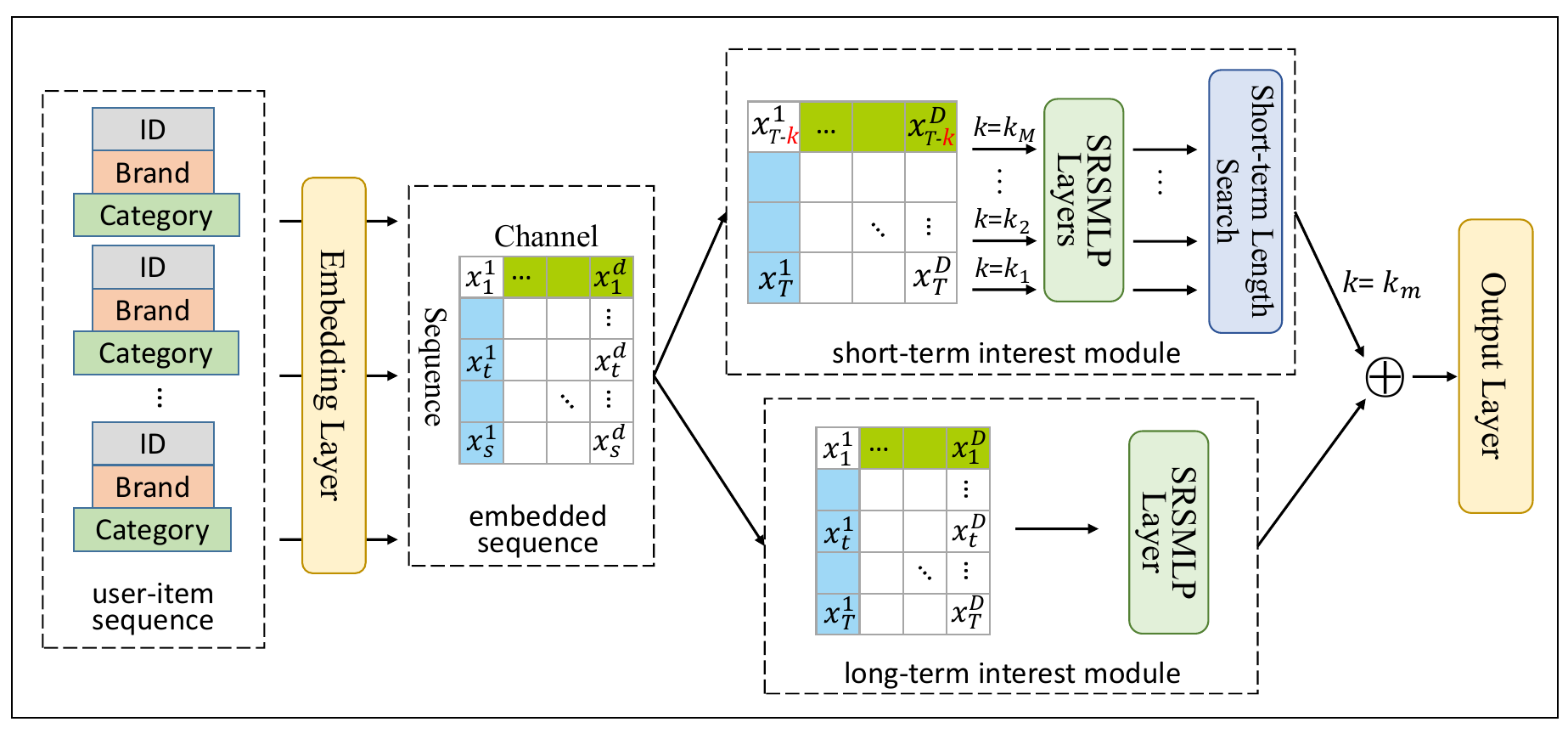}
	\caption{Framework overview of AutoMLP.}
\label{fig:model_framework}
\end{figure*}

Essential information for making informed sequential recommendations could be summarized into three-fold. 
The first fold is the long-term user sequential dependency, which relates to the entire sequence of interacted items. It illustrates the users' relatively static long-term interests.
The second fold is the short-term user sequential dependency, which may deviate from the long-term interest in a dynamic recommendation environment. It describes the users' most recent interest, which could be temporarily independent.
In addition, item features contribute significantly to illustrating the users' short-term dependency~\cite{ma2019hierarchical}, which is the third-fold critical information for sequential recommendations. 
Obviously, users' behaviors in sequential recommendations are usually affected by both their long- and short-term interests, which describe users' preferences from different perspectives, as illustrated in Figure~\ref{fig:Example}. 
However, while the majority of existing works are adept at capturing long-term interest, only a few specifically address the relatively dynamic short-term interest.

Existing methods in sequential recommendation usually tend to model the sequential dependencies among interacted items using Recurrent Neural Network (RNN) ~\cite{hidasi2015session,tan2016improved,wu2017recurrent} and RNN with attention mechanism~\cite{li2017neural,ying2018sequential}. RNN-related methods can successfully capture the short-term dependency, but it is well-known that RNN is prone to degenerate when facing long sequences, even-though techniques such as LSTM and GRU were proposed to mitigate this problem.
Besides, recent advances show that Transformer-based methods can outperform previous methods by a large margin~\cite{kang2018self,sun2019bert4rec,zhao2022mae4rec}. 
However, 
due to the nature of self-attention, they are insensitive to the order of the input sequence, so they must rely on auxiliary positional embedding to learn the sequential information. Despite it has been shown that such an approach sometimes may hurt the performance~\cite{kang2018self,ke2020rethinking}.
In addition, Transformer's computational complexity is quadratic to the input sequence, so the computational cost for handling a long user interaction sequence is not neglectable. 
Furthermore, since users' behaviors could be heterogeneous in the long-term and short-term, \ie the learned short-term dependencies might not accord with the long-term or vice-versa, most existing works train separate models to capture long- and short-term interests, respectively. 
However, they restrict the short-term sequence length through empirical practices, such as selecting a fixed amount of most recent interactions, or from a fixed time interval~\cite{tang2018personalized,ying2018sequential,ma2019hierarchical}. 
These approaches are obviously not adaptive or automated enough, because the optimal short-term length varies across different recommendation tasks or scenarios.

To mitigate aforementioned issues, we propose a simple yet efficient method called \textbf{Auto}mated Long-term Short-term \textbf{M}ulti-\textbf{L}ayer \textbf{P}erceptron for sequential recommendation (\textbf{AutoMLP}). AutoMLP only consists of MLP blocks, therefore maintaining linear time and space complexity. We devise a long-term interest module and a short-term interest module to capture long- and short-term dependencies respectively. To automatically adapt the short-term interest window across different tasks, we leverage continuous relaxation to convert the discrete sequence length to a continuous and differentiable representation via AutoML techniques, which could be optimized via gradient descent ~\cite{liu2018darts}. To show the effectiveness of our proposed method, we test it on commonly used benchmarks. 
To summarize, our paper has the following contributions: 
\begin{itemize}[leftmargin=*]
\item  We propose a new long/short-term sequential recommendation model which only consists of MLP blocks. Despite its simple architecture and linear-complexity, it presents competitive performance against state-of-the-art methods.
\item We devise an automated short-term session length learner to search the local-optimal short-term interest length adaptive to the given context, which enhances the model's generality. 
\item We test our proposed method through extensive experiments on both commonly used datasets and real-world private datasets from the industry. 
\end{itemize}

\section{Framework}

\begin{figure*}[!ht]
	\centering
	\hspace*{-8.1mm}\includegraphics[width=1.08\linewidth]{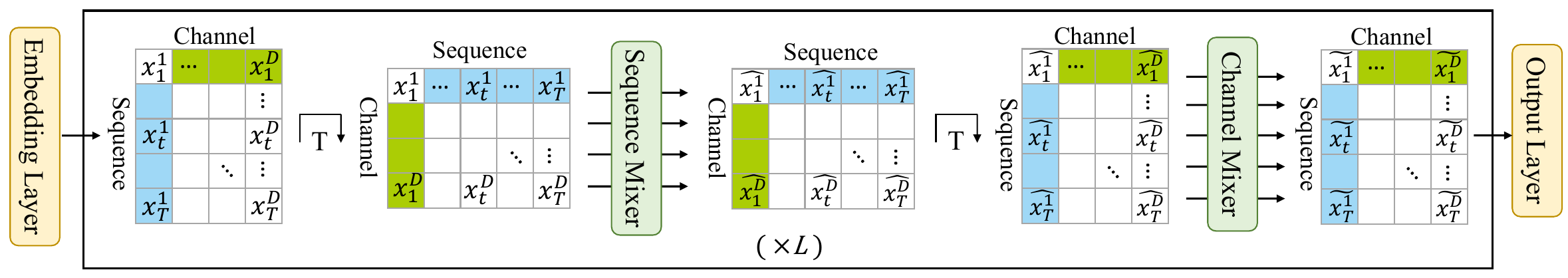}
	\caption{Architecture of Sequential Recommender System MLP Layer (SRSMLP Layer). }
\label{fig:model_SRSMLP}
\vspace{-3mm}
\end{figure*}

In this section, we will make a detailed introduction to our proposed method - AutoMLP, which can adaptively choose users' short-term interest length, and use MLP blocks to capture long-term and short-term interests efficiently. We will first present the overview of the AutoMLP framework, and then discuss the components in detail. 

\subsection{Problem Formulation}

The goal of sequential recommendation is to predict the user's next possible interacted item given their historical interaction sequences. Thus, for each user $u$ in the user set $U$, there is a corresponding sequence from \(S = \{S_{1},...,S_{u},...,S_{U}\}\), where \(S_{u}\) indicates the interaction sequence for user \(u\). Within each sequence, user-interacted items will be sorted chronologically, \ie \(S_{u} = \{i_{1},...,i_{t},...,i_{T}\}\),
where $i_t$ is the item interacted at timestamp $t$, and $T$ is the sequence length.
Then, the objective of sequential recommendation can be formally defined as: \textit{given user \(u\)'s historical interaction sequence \(S_{u}\), the goal is to find a function $f: S_{u} \rightarrow i_{T+1}$, where $i_{T+1}$ is the next item to be recommended to the user \(u\)}.

\subsection{Framework Overview}

Now we introduce the overview of AutoMLP - a sequential recommender system based solely on MLP architectures, which can automatically learn the appropriate short-term user interest length for different sequential recommendation applications. The main body of AutoMLP is made up of two separate MLP-based networks, namely the long-term user interest module and short-term interest module as shown in Figure \ref{fig:model_framework}. To be specific, the long-term user interest module takes the entire user historical behavior sequence for prediction and therefore is more leaning towards long-term sequential dependencies.

On the other hand, the short-term user interest module takes a certain number of latest interactions before time $T$, tending to model the short-term sequential dependencies. 
The number of recent interactions $k$ will be determined by a Neural Architecture Search (NAS) algorithm, DARTS~\cite{liu2018darts}, which leverages continuous relaxation to make the neural architecture search space differentiable and thus can be optimized by gradient descent.
Finally, the outputs of separated modules will be fused by a fully-connected layer to predict the next item interacted.

\subsection{Detailed Architecture}

In this section, we will introduce the macro-structures as well as the methodology behind our proposed method in detail. 

\subsubsection{\textbf{Embedding Layer}}

Following the commonly used strategy in sequential recommender system~\cite{kang2018self,zhang2019feature,sun2019bert4rec}, we use a lookup table to transform the item ID and other features into embedding vectors. Assuming each item $i_t$ has $C$ features, we then use a fully-connected layer to fuse $C$ embedding vectors from different features into one embedding vector $\boldsymbol{x_t} = [x_t^1,\cdots,x_t^D]$ with predefined dimension $D$. For a user interaction sequence with length $T$, we then attain an embedding table with size \(T \times D\), \ie the embedded sequence in Figure \ref{fig:model_framework}.

\subsubsection{\textbf{Long-term Interest Module}}

Now we present the key component for modeling the long-term sequential dependencies, \ie long-term interest module. As we mentioned above, the long-term interest module takes the entire user interaction sequence as input and encoded the interest as the hidden representation for prediction. The encoding process is carried out by a sequential recommender system MLP layer, \ie \textit{SRSMLP Layer} in Figure \ref{fig:model_SRSMLP}, which is comprised of two MLP blocks, namely the \textit{sequence-mixer} and \textit{channel-mixer}. SRSMLP Layer alternately employs both MLP blocks to capture the sequential correlations within the user interaction sequence and the cross-channel correlations within each item's embedding vector.

\begin{figure}[t]
	\centering
	\includegraphics[scale=0.266]{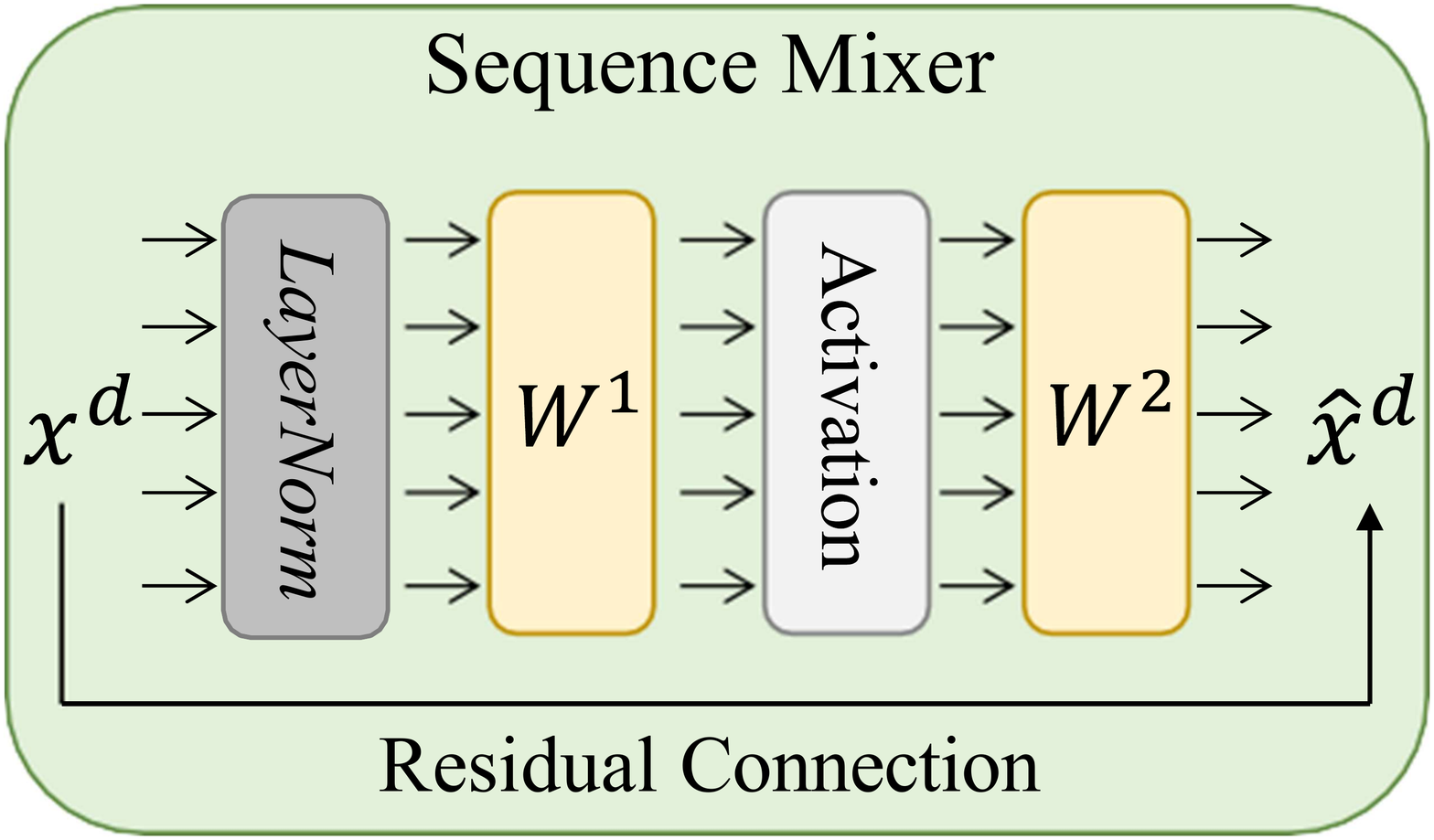}
	\caption{Sequence mixer. Channel mixer has a similar architecture, except for the input $x^d$ and output $\hat{x}^d$.}
\label{fig:model_sequence}
\vspace{-5mm}
\end{figure}

\textit{\textbf{Sequence-mixer.}} The illustration of the structure of the sequence-mixer is shown in Figure \ref{fig:model_sequence}. The objective of the sequence-mixer is to learn the sequential information within the user interaction sequence. The input dimension of the sequence-mixer is the length of the input sequence, which is $T$. Recall that after the embedding layer, we have an embedding table with a size of \(T \times D\), so sequence-mixer is applied to the transposed embedding table with a size of \(D \times T\), as shown in Figure \ref{fig:model_SRSMLP}. Sequence-mixer performs non-linear projection that maps input $\boldsymbol{x^d}$ to output $\boldsymbol{\hat{x}^d}$ of the same shapes, \ie \(\mathbb{R}^{D \times T} \mapsto \mathbb{R}^{D \times T}\), fusing the cross-item (sequential) information within the sequence to each item's embedding vector. More specifically, since the inputs are the same embedding dimension of the entire sequence, \ie $|[x_1^d,\cdots,x_T^d]|=T, \forall d \in [1,D]$, we can observe that sequence-mixer can learn the sequential correlation throughout the sequence, which often reveals users' long-term interest evolution. We can denote the output of the sequence-mixer at layer \(l\) as:
\begin{equation} 
 \boldsymbol{\hat{x}^d} = \boldsymbol{x^d} + \boldsymbol{W^2} g^l(\boldsymbol{W^1} LayerNorm(\boldsymbol{x^d}))
\end{equation}
where \(d = 1, 2, \cdots, D\). \(\boldsymbol{x}^d\) is the $d$th example from the embedding table . \(\boldsymbol{\hat{x}^d}\) is the output of the sequence-mixer block, \(g^l\) is the non-linear activation function at layer \(l\),  \(\boldsymbol{W^1} \in \mathbb{R}^{R_s \times T}\) denotes the learnable weights representing the first fully connected layer in the sequence-mixer,  \(\boldsymbol{W^2} \in \mathbb{R}^{T\times R_s}\) signifies the learnable weights of the second fully connected layer in the sequence-mixer, \(R_s\) is the tunable hidden size of sequence-mixer. We employ layer normalization (\(LayerNorm\)) and residual connection as in MLP-mixer~\cite{tolstikhin2021mlp}.

\textit{\textbf{Channel-mixer.}} Sharing a similar macro-structure to sequence-mixer in Figure \ref{fig:model_sequence}, the key distinction between channel-mixer and sequence-mixer is their purpose. The objective of the channel-mixer is to learn the correlation within each item's embedding vector. Since each dimension of an embedding vector usually expresses different semantics, collectively learning their representation is vital for making an informed prediction~\cite{lee2021moi}. Also, after the sequence-mixer, the sequential information is fused in the individual embedding dimension of each embedding vector, channel-mixer can also communicate the sequential information from different dimensions and thus coherently learn the hidden representation of sequential information~\cite{tolstikhin2021mlp}. Therefore, the input dimension of the channel-mixer must match the size of the embedding vector, which is $D$. So for the embedding table, we obtained from the embedding layer, after passing through the sequence-mixer and encoding with sequential information, we first transpose it back to the initial shape, \ie \(T \times D\). Then, channel-mixer maps input $\boldsymbol{x^t}$ to output $\boldsymbol{\hat{x}^t}$ of the same shapes, \ie \(\mathbb{R}^{T \times D} \mapsto \mathbb{R}^{T \times D}\), fusing the cross-channel correlation to each item's embedding vector. We can denote the output of the channel-mixer at layer \(l\) as:
\begin{equation} 
 \boldsymbol{\hat{x}^t} = \boldsymbol{x^t} + \boldsymbol{W^4} g^l(\boldsymbol{W^3} LayerNorm(\boldsymbol{x^t}))
\end{equation}
where \(t = 1, 2, \cdots, T\), $\boldsymbol{x^t}$ is the input, which is the $t^{th}$ example from the embedding table. $\boldsymbol{\hat{x}^t}$ is the output of the channel-mixer that contains cross-channel correlations. \(\boldsymbol{W^3} \in \mathbb{R}^{R_{c}\times D}\) is learnable weights of the first fully connected layer in the channel-mixer, and \(\boldsymbol{W^4} \in \mathbb{R}^{D\times R_{c}}\) is that of the second fully connected layer, where \(R_{c}\) denotes the tunable size of hidden layer in channel-mixer.

\begin{figure}[t]
	\centering
	\includegraphics[scale=0.72]{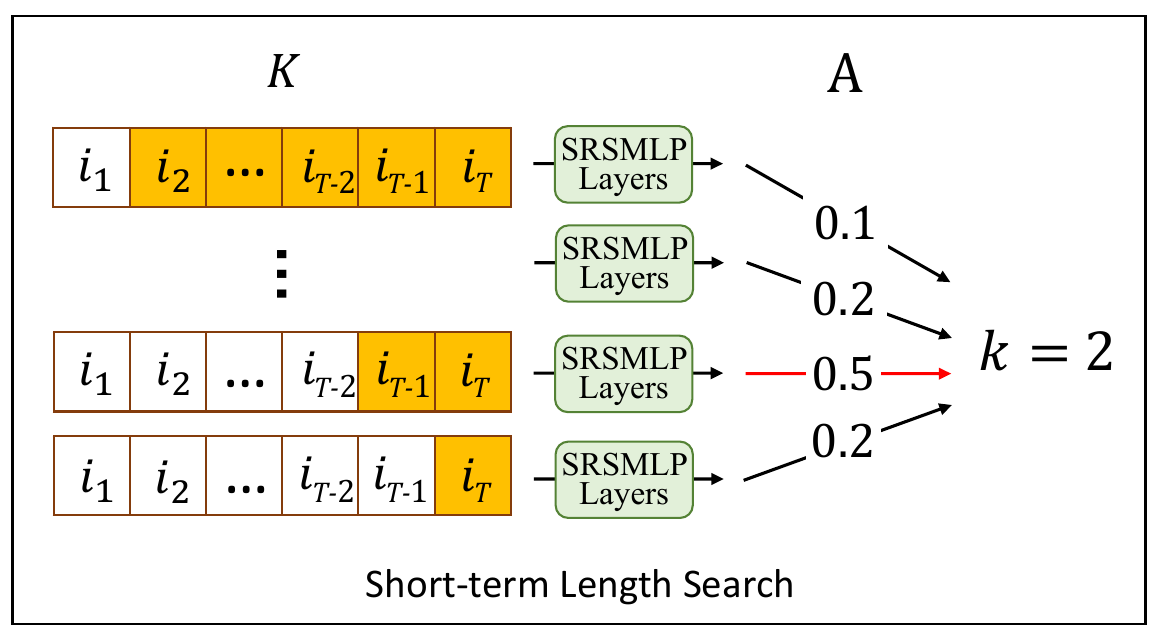}
	\caption{Short-term session length search process. $K$ is the set of candidate lengths, represented by the highlight part.}
\label{fig:model_nas}
\end{figure}

\subsubsection{\textbf{Short-term Interest Module.}}

Now we introduce the short-term interest module for capturing the short-term sequential dependencies in users' historical sequences. The short-term interest module follows a similar framework as the long-term interest module, meaning it also contains a sequence-mixer and channel-mixer. The key difference is the short-term interest length search section that can automatically select optimal short-term interest length $k$ in a data-driven manner. 

\textit{\textbf{Session Length Search.}} The session length search part is a Neural Architecture Search (NAS) algorithm based on DARTS~\cite{liu2018darts}. To be specific, we first define some candidate short-term lengths before the last item $i_T$, representing the possible range of users' short-term interest length, \ie \(\boldsymbol{K} = \{k_{1},...,k_{m},...,k_{M}\}\), as shown in Figure \ref{fig:model_nas}. Then we create $M$ SRSMLPs, each for individual candidate $k_m$. Next, as shown in Figure \ref{fig:model_nas}, we assign a set of learnable \textit{architectural weights} \(\boldsymbol{A} = \{\alpha_{1},...,\alpha_{m},...,\alpha_{M}\}\) to the outputs from different SRSMLPs respectively. After that, we apply softmax to transform weights as continuous and differentiable approximations:

\begin{equation} 
\label{softmax}
p_m = \frac{ \exp(\alpha_m)} {\sum_{m=1}^{M} \exp(\alpha_j)}
\end{equation}
where $m = 1,\cdots,M$, and \(\alpha_m\) is the learnable weight for $m^{th}$ candidate short-term sequence. After Equation \eqref{softmax}, we now have a set of continuous and differential weights $\{p_1,\cdots,p_m,\cdots,p_M\}$ that could adaptively decide the length of the short-term sequence.

\textbf{\textit{Discussion.}} Since the effect between the value of short-term interest length and the model performance is not monotonic, thus to determine a local optimal value must apply an exhaustive search, which is extremely computationally expensive for a long user-item interaction sequence, since there are more possible candidates. Therefore, the main upside of such an approach is to learn a local optimal user short-term interest length without enumerating every possible model architecture and train them repetitively, thus making the decision process of selecting short-term interest length highly efficient and adaptive. 

\subsubsection{\textbf{Output Layer}}

Once we find the optimal length of the short-term sequence, we can obtain the outputs from the long-term interest module and the optimal short-term interest module. Then, we use a fully-connected layer to learn their joint representation as follows:

\begin{equation} 
\begin{aligned}
 \boldsymbol{h_T} = \boldsymbol{W^o} LayerNorm(\boldsymbol{x_T^s;x_T^l})) + \boldsymbol{b^o}
 \label{connect}
 \end{aligned}
\end{equation}
where $h_T$ is the final hidden representation of time step $T$, $x_T^s$ and $x_T^l$ are the output from the short-term interest module and long-term interest module at time step $T$, respectively. \(\boldsymbol{W^o}\) is learnable weights that projects the combined $x_T^s$ and $x_T^l$ to a lower dimension representation, hence \(\boldsymbol{W^o} \in \mathbb{R}^{D\times 2D}\), and \(\boldsymbol{b^o} \in \mathbb{R}^{D}\) is learnable bias vector. Finally, $h_T$ will be used for predicting users' next interaction, which we will later introduce in Section \ref{sec:inference} of \textit{model inference}. 

\subsection{Training and Inference}

Following the common setting in the sequential recommendation, our training loss is defined by a Cross-Entropy loss function: 
\begin{equation} 
 \mathcal{L} = -\sum_{S_u \in S}\sum_{t \in [1,...,T]}[log(\sigma({r_{i_t,t}})) + \sum_{j \not\in S_n}log(1-\sigma(r_{i_j,t}))]
\end{equation}

where \(\sigma\) demotes sigmoid function. \({r_{i_t,t}}\) is model's predicted similarity to ground-truth item \(i_t\), and \(r_{i_j,t}\) is the predicted similarity to sampled items at time step \(t\), where \(j\) is the negative sampled items. \(S\) is the superset that contains all users' interaction sequences. 

\subsubsection{\textbf{Training}} The training process contains two phases. The first phase is the search phase, aiming to find local optimal \(\boldsymbol{A}^{*}\) representing preferable short-term length. The second phase is the retraining phase, where after \(\boldsymbol{A}^{*}\) is found, we retrain the AutoMLP framework with the optimal short-term length.

\textbf{\textit{Search Phase}}.
The previous section illustrates how continuous relaxation can make the search space \(\boldsymbol{K}\) differentiable and thus could be optimized via back-propagation. 
Now our objective is to jointly learn the architectural weights \(\boldsymbol{A} = \{\alpha_m\}\), and all other learnable parameters of AutoMLP \(\boldsymbol{W}\). Note that each $p_m$ is corresponding with each $\alpha_m$, \ie learning $\{p_m\}$ is the same to learning $\{\alpha_m\}$.
While \(\boldsymbol{A}\) is a subset of the learnable parameters of AutoMLP, literature shows that simply updating \(\boldsymbol{W}\) and \(\boldsymbol{A}\) altogether will cause overfitting problems in the training process, since they are highly dependent from each other ~\cite{zhao2021autoloss,zhao2021autodim}. 
Therefore, we will use the training dataset to optimize \(\boldsymbol{W}\) following common practice, while using the validation dataset to optimize \(\boldsymbol{A}\). To be specific, we formulate this as a bilevel optimization~\cite{colson2007overview}, with \(\boldsymbol{A}\) as upper-level variable and \(\boldsymbol{W}\) as lower-level variable ~\cite{liu2018darts}. Formally, we denote that as:

\begin{equation} 
\begin{aligned}
\label{argmin}
\min_{\boldsymbol{A}} \; &\mathcal{L}_{val} \big(\boldsymbol{W}^*(\boldsymbol{A}),\boldsymbol{A}\big)\\
s.t. \; & \boldsymbol{W}^*(\boldsymbol{A}) = \arg\min_{\boldsymbol{W}} \mathcal{L}_{train} (\boldsymbol{W}, \boldsymbol{A})
\end{aligned}
\end{equation}

In addition, it is notable that the inner optimization (lower-level variable optimization) for solving Equation (\ref{argmin}) is usually computationally expensive, since every time we intend to optimize \(\boldsymbol{A}\), we have to first train \(\boldsymbol{W}\) till converged. Alternatively, we apply a one-step approximation: 
\begin{equation}
\begin{aligned}
\label{equ:approximation}
& \boldsymbol{W}^*(\boldsymbol{A})\approx \boldsymbol{W} - \xi \nabla_{\boldsymbol{W}}\mathcal{L}_{train} (\boldsymbol{W}, \boldsymbol{A})
\end{aligned}
\end{equation}
where \(\xi\) is the learning rate. The motivation of this approximation is to approximate \(\boldsymbol{W}^*(\boldsymbol{A})\) by a single training step instead of from training thoroughly. The detailed training algorithm of the search phase is shown in Algorithm \ref{alg:opt}. 

\begin{algorithm}[t]
	\caption{\label{alg:opt} Optimization for AutoMLP in Search Phase}
	\raggedright
	{\bf Input}: embedding table $\boldsymbol{x}$ and ground-truth interaction sequences $\boldsymbol{S}$\\
	{\bf Output}: well-learned parameters $\boldsymbol{W}^*$ and $\boldsymbol{A}$\\
	\begin{algorithmic} [1]
		\WHILE{not converged}
		\STATE Sample a mini-batch of validation data examples 
		\STATE Estimate the approximation of $\boldsymbol{W}^*(\boldsymbol{A})$ via Eq.(\ref{equ:approximation})
		\STATE Update $\boldsymbol{V}$ by descending $\nabla_{\boldsymbol{A}} \;\mathcal{L}_{val} \big(\boldsymbol{W}^*(\boldsymbol{A}),\boldsymbol{A}\big)$
		\STATE Sample a mini-batch of training data examples
		\STATE Update $\boldsymbol{W}$ by descending $\nabla_{\boldsymbol{W}}\mathcal{L}_{train} (\boldsymbol{W}, \boldsymbol{A})$
		\ENDWHILE
	\end{algorithmic}
\end{algorithm}

\textbf{\textit{Retraining Phase}.} In the search stage, we incorporate all candidate short-term sequences into the AutoMLP framework, where the suboptimal short-term sequences may lead to a negative impact on model performance. Thus, we need to retrain AutoMLP with only the optimal short-term sequence.

To be specific, after the optimization of the search stage, we obtain well-trained \(\boldsymbol{A}^{*}\). We select the optimal short-term length with the highest $\alpha_m$, and discard all other candidate lengths. Then, we retrain the AutoMLP framework based on the outputs from the long-term interest module and the optimal short-term interest module following common deep recommender system training paradigm, \ie optimizing loss $\min_{\boldsymbol{W}} \mathcal{L}_{train} (\boldsymbol{W})$ by gradient descent.

\subsubsection{\textbf{Inference}}
\label{sec:inference}
Our proposed method employs a commonly used inference process in sequential recommender systems~\cite{kang2018self,sun2019bert4rec,zhang2019feature,zhou2020s3}, which is to compute the cosine similarity between model output and all candidate items. More specifically, once we have the hidden representation $h_T$ from the output layer as the final hidden representation of the entire sequence, we then calculate the dot product of $h_T$ and the item embeddings of all candidate items $I$:
\begin{equation} 
    p_{\iota} = \text{softmax}(h_T \cdot E_\iota^T)
\end{equation}
where $\iota = 1,2,\cdots,|I|$, and $|I|$ is the total number of candidate items. \(E_\iota\) is the item embedding of \(i_\iota\), \(p_{\iota}\) is the predicted probability for \(i_{\iota}\) being the next possible interaction. Then, candidate item with highest \(p_{\iota}\) is the predicted next interaction \(i_{T+1}\). 

\subsection{Complexity Analysis}

This section will show the important strength of AutoMLP, which is the linear computational complexity. The complexity of AutoMLP is primarily determined by two MLP blocks, sequence-mixer, and channel-mixer. Since the time complexity of an MLP with one hidden layer is $O(input$ $units$ $\times$ $hidden$ $units$ $\times$ $output$ $units)$, thus we can know that the complexity of sequence-mixer is $O(T \times R_s + R_s \times T)$. Since the determinant variable here is maximum sequence length $T$, and $R_s$ is a constant that is independent of $T$, we can rewrite that as $O(T)$. Similarly, we have the complexity of the channel-mixer as $O(D)$. And since short-term modules have the exact same structure except for the sequence length, we have the complexity of the short-term sequence-mixer as $O(k)$, and the complexity of the short-term channel-mixer as $O(D)$ as well. Hence the complexity of AutoMLP is $O(T + D + k)$, and it is linear to the three determinant variables.

\begin{table}[]
\begin{center}
	\caption{Statistics of the datasets.}
	\label{table:statistics}
	\scalebox{1.1}{
	\begin{tabular}{@{}|c|c|c|@{}}
		\toprule[1pt]
		Data & MovieLens & Amazon Beauty \\ \midrule
		\# Interactions  & 1,000,209 & 2,023,070 \\
		\# Items  & 3,952 & 249,274 \\
		\# Users  & 6,040 & 1,210,271 \\
		\bottomrule[1pt]
	\end{tabular}}
 \vspace{-5mm}
\end{center}
\end{table}

\begin{table*}

\centering
\caption{Overall performance comparison. Best performances are bold, next best performances are underlined}
 \vspace{-3mm}
\scalebox{1.08}{
\renewcommand{\arraystretch}{1.2}
\begin{tabular}{c||ccc||ccc||c} 
\hline
\multicolumn{1}{c||}{Methods} & \multicolumn{3}{c||}{MovieLens} & \multicolumn{3}{c||}{Beauty} & \multirow{2}{*}{Param}\\ 

\multicolumn{1}{c||}{Metrics} & MRR@10 & NDCG@10 & HR@10 & MRR@10 & NDCG@10 & HR@10\\
\hline
\multicolumn{1}{c||}{FPMC} & 0.2453 & 0.3088 & 0.5156 & 0.0991 & 0.1251 & 0.2098 & 100 M \\
\multicolumn{1}{c||}{GRU4Rec}  & \underline{0.3893} & \underline{0.4553} & 0.6666 & 0.1162 & 0.1435 & 0.2324 & 33.5 M\\
\multicolumn{1}{c||}{BERT4Rec}  & 0.3535 & 0.4289 & \underline{0.6695} &  0.0907& 0.1198 & 0.2154 & 17 M   \\
\multicolumn{1}{c||}{NextItNet}  & 0.2085 & 0.2642 & 0.4455 & 0.1087 & 0.1393 & 0.2385 & 16.8 M\\ 
\multicolumn{1}{c||}{GRU4Rec$^+$}  & 0.3736 & 0.4412 & 0.6578 & \underline{0.1325} & \underline{0.1638} & \underline{0.2657} & 34.1 M \\
\multicolumn{1}{c||}{FDSA}  & 0.3725 & 0.4409 & 0.6594 & 0.1305 & 0.1595 & 0.2536 & 34.3 M \\ 
\multicolumn{1}{c||}{\textbf{AutoMLP}}  & \textbf{0.3912*} & \textbf{0.4593*} & \textbf{0.6767*} & \textbf{0.1438*} & \textbf{0.1754*} & \textbf{0.2779*} & 16.8 M \\
\hline
\end{tabular}
\label{table:overall}
}
\\``{\large *}'' indicates the statistically significant improvements (i.e., two-sided t-test with $p < 0.05$) over the original model. 
\\``Param'' refers to the number of trainable model parameters on Beauty dataset, and M = million.

\end{table*}

\section{Experiments}

In this section, we will evaluate our proposed method - AutoMLP through extensive experiments on two commonly used benchmark datasets in the recommender system. 

\subsection{Datasets}

\textbf{MovieLens}\footnote{https://grouplens.org/datasets/movielens/1m/}: MovieLens is a commonly used benchmark dataset for recommender systems, it contains users' rating history on a wide range of movies with corresponding timestamps. In this study we choose MoiveLens-1M for experiments, which contains over a million interactions, detailed statistics are available in Table \ref{table:statistics}. 

\textbf{Amazon Beauty}\footnote{http://jmcauley.ucsd.edu/data/amazon/}: This dataset contains reviews and ratings of various items that are crawled from Amazon ~\cite{mcauley2015image}, it is divided into multiple categories, and in this paper, we will use "Beauty" category, statistics of this category are available in Table \ref{table:statistics}.

In order to model the long-term dependencies in a more realistic scenario, we filter out users who have less than 10 interactions in MovieLens and Amazon Beauty.
More statistics about the datasets can be found in Table~\ref{table:statistics}.

\subsection{Experimental Setting}

We follow the common scenario in the sequential recommendation, which is the next-item prediction (leave-one-out evaluation). In this setting, we use the last interactions from user interaction sequences as the test set, the second last interactions as the validation set, and all of the previous interactions as training set ~\cite{sun2019bert4rec,kang2018self,zhang2019feature,zhou2020s3,li2020time}. We will also pair 100 negative samples for each ground-truth item in evaluation, sampling based on their popularity ~\cite{sun2019bert4rec,huang2018improving}. 

\subsection{Evaluation Metrics}
We adopt the three most commonly used metrics for evaluation, namely Hit Ratio (HR), Normalized Discounted Cumulative Gain(NDCG), and Mean Reciprocal Rank (MRR). HR is the accuracy for the ground-truth item appearing in top $N$ recommendations, NDCG is a ranking loss that measures the position of the ground-truth item in the top $N$ recommendations, and MRR is the reciprocal of the ground-truth item’s ranking in top $N$ recommendations.

\subsection{Baselines}

To illustrate the effectiveness of our proposed method, we compare it against several popular sequential recommendation models, where some of them achieve state-of-the-art performances. Of their design motivation and adapted techniques, they could be divided into three groups:

\noindent\textbf{General sequential recommender systems} refers to the representative sequential recommendation models of some important categories, namely MC-based, RNN-based, and Transformer-based. 

\begin{itemize}[leftmargin=*]

\item \textbf{FPMC} Factorizing Personalized Markov Chains (FPMC) \cite{rendle2010factorizing} represents the earlier works in sequential recommender systems, which is using Markov Chains to model the sequential dependencies between items. 

\item \textbf{GRU4Rec} Hidasi \etal proposed using RNN for the sequential recommendation, and improved vanilla RNN with Gated Recurrent Unit(GRU) and point-wise ranking loss~\cite{hidasi2015session}. 

\item \textbf{BERT4Rec} Sun \etal utilized Bi-directional self-attention to improve transformer-based sequential recommendation~\cite{sun2019bert4rec}, and several studies have shown that BERT4Rec surpasses vanilla self-attention approach such as SASRec constantly~\cite{sun2019bert4rec,lee2021moi,zhou2020s3}

\end{itemize} 

\noindent\textbf{Long-term short-term sequential recommender systems} focus on modeling the short-term sequential dependencies while maintaining long-term sequential dependencies as well: 

\begin{itemize}[leftmargin=*]

\item \textbf{NextItNet} Yuan \etal pointed out some existing limitations of Caser~\cite{tang2018personalized}, especially its drawback in modeling long-term sequential dependencies and proposed a generative model which does not require pooling operation for a bigger receptive field to improve CNN-based sequential recommender systems~\cite{yuan2019simple}. 

\end{itemize}

\noindent\textbf{Feature-level sequential recommender systems} refer to the sequential recommendation models that can learn information within item features to aid next-item prediction. 

\begin{itemize}[leftmargin=*]

\item \textbf{GRU4Rec$^+$} is the improved version of GRU4Rec, which leverages item features for better prediction under a feature-rich scenario and parallel RNN for higher efficiency~\cite{hidasi2016parallel}.

\item \textbf{FDSA} Feature-level Deeper Self-Attention Network for Sequential Recommendation (FDSA)~\cite{zhang2019feature} utilize item features for the sequential recommendation, and has been considered to have state-of-the-art performance in supervised-methods~\cite{zhou2020s3}.

\end{itemize}

\subsection{Implementation Details}

The implementations of baseline methods as well as AutoMLP are based on RecBole framework ~\cite{zhao2021recbole}, an open-sourced library for recommender systems, which offers an unbiased environment to evaluate the performance of our proposed methods. 

For baseline methods, we use hyper-parameters as suggested in the original papers. For those unspecified hyper-parameters, we use grid search for hyper-parameter tuning. The searching strategy is to select the hyper-parameters that yield the best performance on the validation set. We use the early-stop strategy, \ie if, in the next 10 epochs, the validation performance does not improve, we stop the training and use the model from the current best validation performance. For some larger models such as BERT4Rec, when dealing with large datasets, their space complexity could be too large to fit into GPU memory, therefore we define a "space efficient" setting, referring to the search best hyper-parameters within our GPU memory capacity, which is 32GB. We use Adam~\cite{kingma2014adam} as the optimizer for all implementations, the learning rate is set as $1e-3$, \(\beta_1\) as 0.9, \(\beta_2\) as 0.999.


\subsection{Overall Performance}
In Table \ref{table:overall}, we show the overall comparison of our proposed methods against a wide range of popular baseline methods. 

From the results, we can make the following observations: 
\begin{itemize}[leftmargin=*]
\item When item features are available, feature-level models usually obtain better performances than those who cannot process features, indicating the importance of item features in the sequential recommendation. 
\item AutoMLP maintains competitive performance across all datasets. Specifically, AutoMLP exceeds existing long/short-term sequential recommender systems such as NextItNet significantly, showing it is not only an efficient choice but also a strong alternative in terms of accuracy. 
\item Finally, we can observe that AutoMLP consumes the least trainable model parameters, which indicates its superior space efficiency in real-world sequential recommender systems. 
\end{itemize}
In summary, AutoMLP can achieve comparable performances against popular baselines on public datasets with lower space complexity, which validates its effectiveness. 

\subsection{Efficiency Analysis}
In this section, we will study the time/space efficiency of AutoMLP, which are critical metrics to launch a recommendation model in industrial recommender systems.

\begin{figure}[t]
	\includegraphics[scale=0.25]{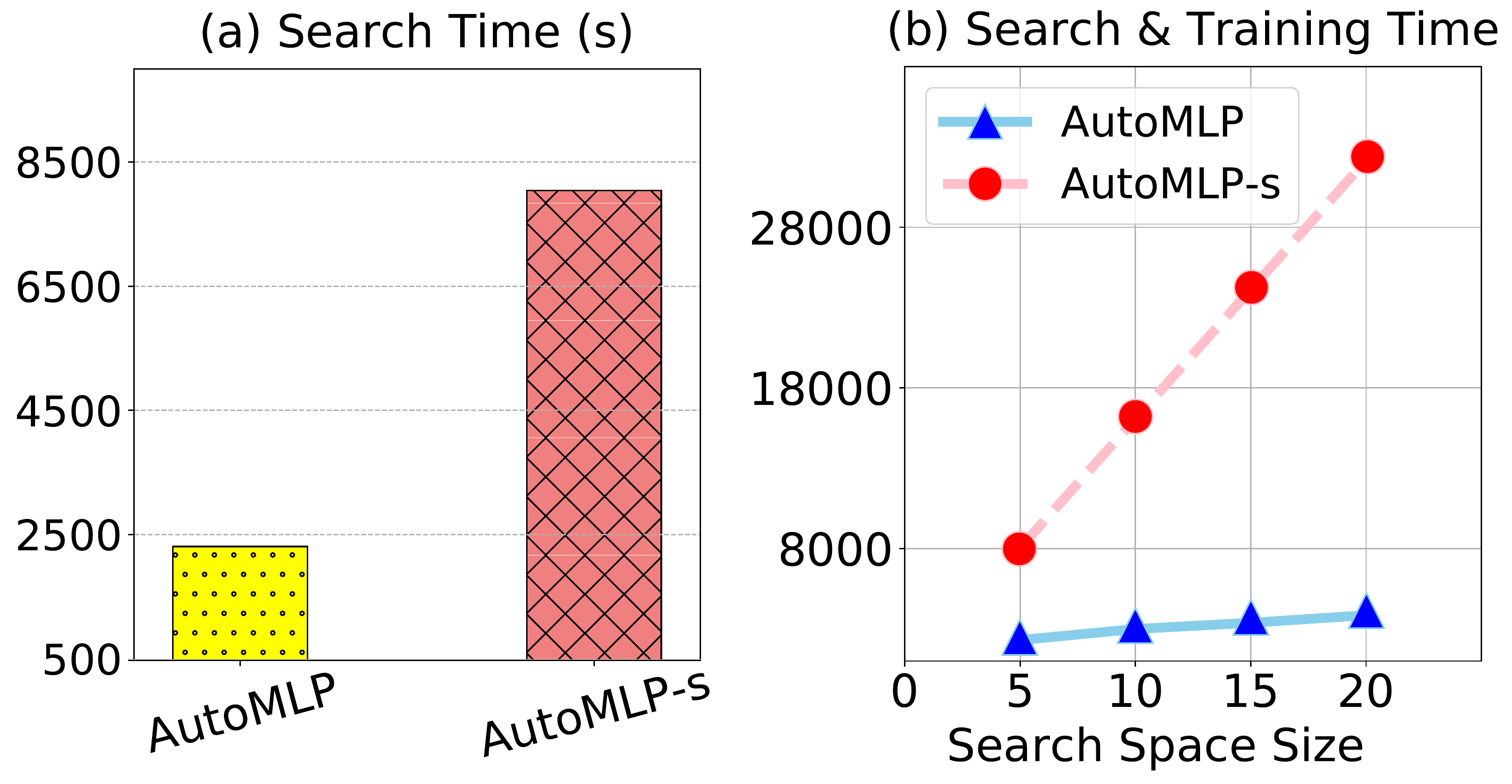}
    \vspace{-5mm}
    \caption{Searching efficiency on Beauty dataset}
    \label{fig:search}
    \vspace{-4mm}
\end{figure}

As one of the fundamental motivations, we argue that via automated short-term interest length search, AutoMLP is more efficient than the exhaustive search for optimal short-term interest length. In Figure \ref{fig:search}, we compare AutoMLP against AutoMLP-s, which has identical architecture as AutoMLP but a different short-term interest length search algorithm. Specifically, AutoMLP-s exhaustively searches for a short-term interest length that yields the best performance as existing methods~\cite{tang2018personalized, ying2018sequential}. In Figure \ref{fig:search}(a), we compare the search time of AutoMLP and AutoMLP-s when the size of candidate search space (\ie candidate short-term length) is 5. And in Figure \ref{fig:search}(b), we show that as the size of the search space increases, the increment ratio of AutoMLP is much smaller, suggesting better scalability of AutoMLP.

\begin{figure}[t]
     \centering
    {\subfigure{\includegraphics[width=0.490\linewidth]{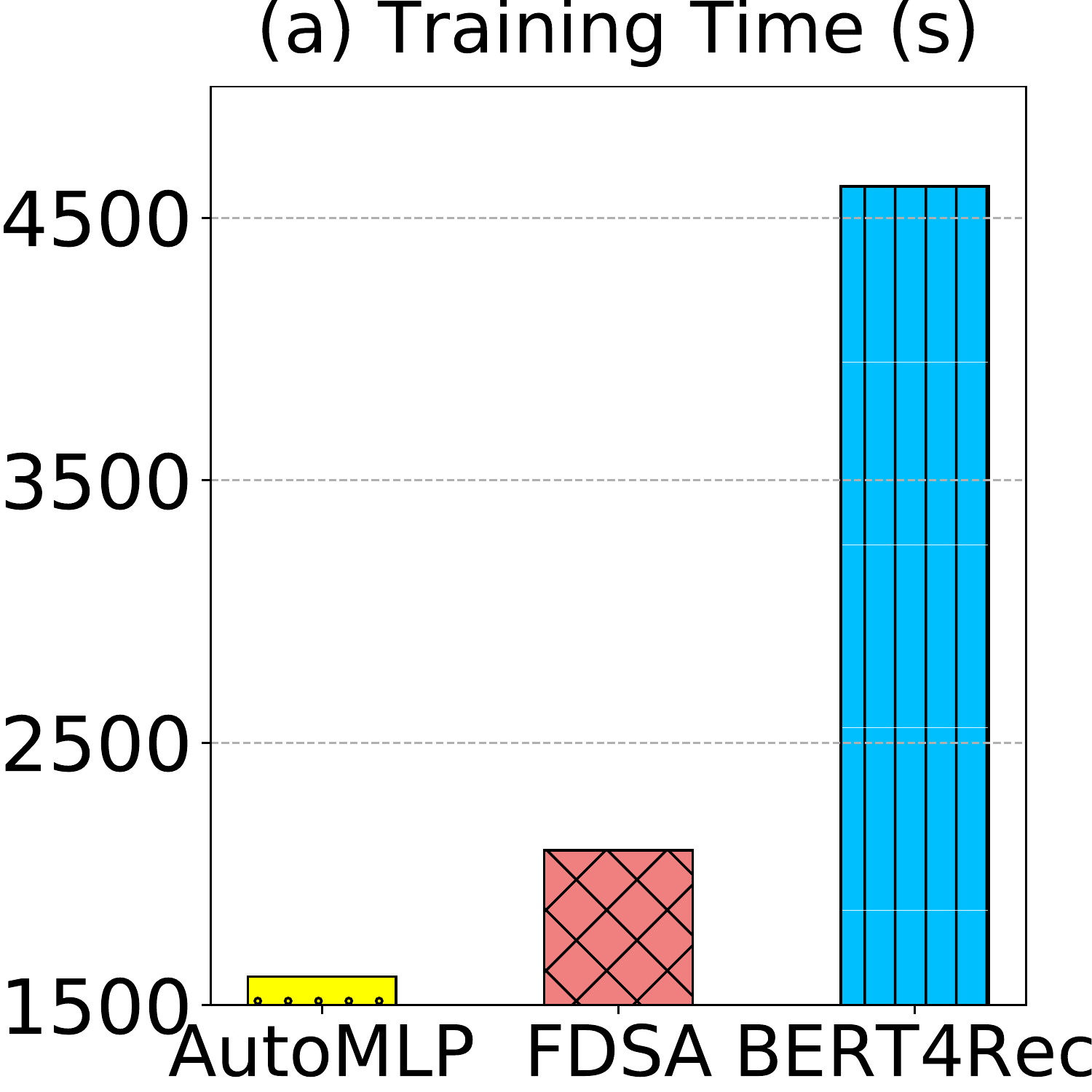}}}
    {\subfigure{\includegraphics[width=0.490\linewidth]{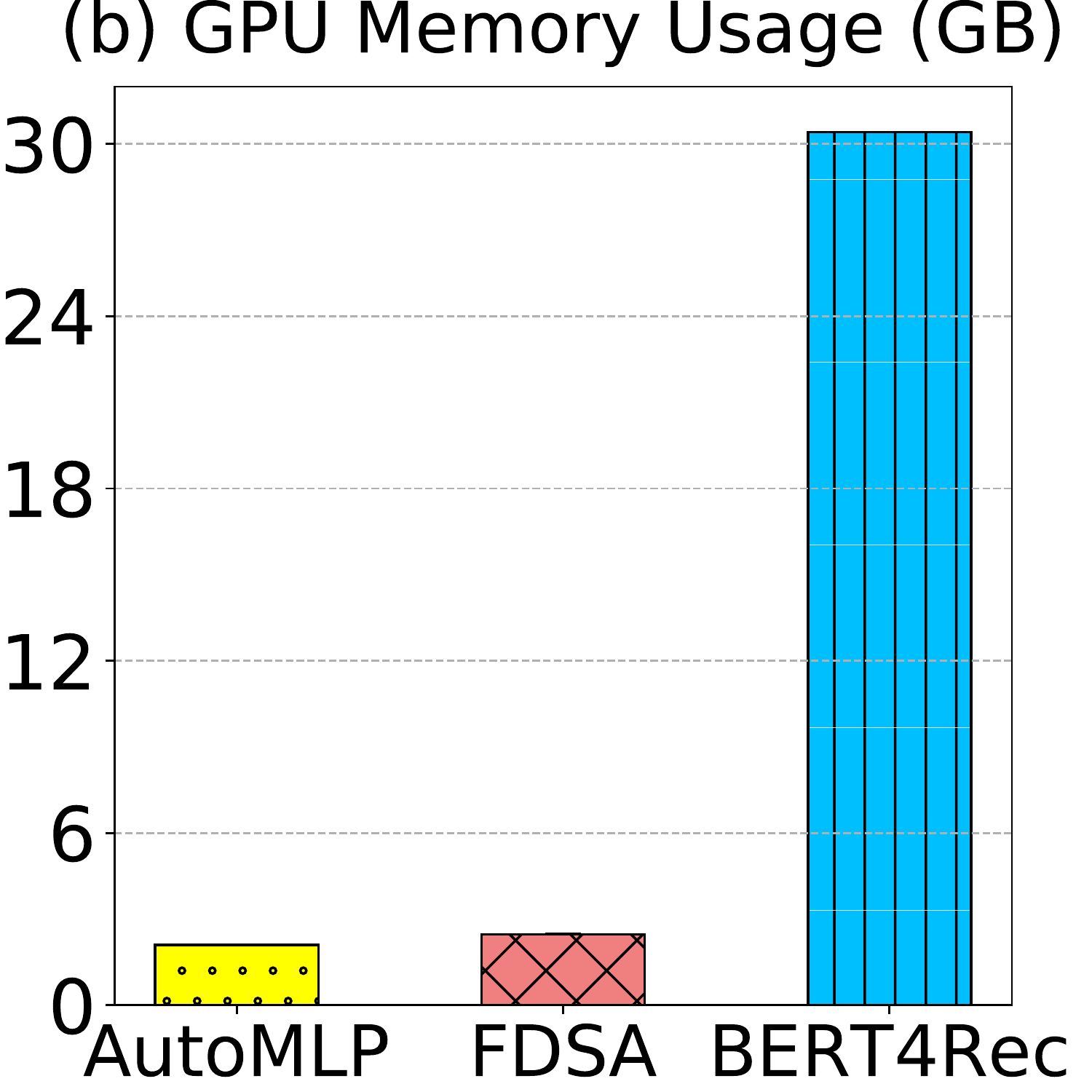}}}
    \vspace{-4mm}
    \caption{Training efficiency on Beauty dataset}
    \label{fig:efficiency}
    \vspace{-4mm}
\end{figure}

In addition, from Figure \ref{fig:efficiency}, we can also observe that AutoMLP shows higher training efficiency in terms of time usage and memory consumption against state-of-the-art Transformer-based methods. Figure \ref{fig:efficiency}(a) shows the total training time used for training AutoMLP, FDSA, and BERT4Rec till convergence in seconds. We can observe that, because BERT4Rec implements complex training techniques such as Cloze objective and bi-directional self-attention~\cite{sun2019bert4rec}, its training efficiency is relatively low. While FDSA obtained a more balanced training efficiency by combining self-attention and vanilla attention, its training cost is still more expensive than AutoMLP. 


\subsection{Hyper-parameters Analysis}
In this section, we present how the essential hyper-parameters in AutoMLP affect the model performance. Unlike transformer-based methods~\cite{kang2018self,sun2019bert4rec,zhang2019feature} and CNN-based methods~\cite{tang2018personalized,yuan2019simple}, which have bigger sets of hyper-parameters and larger search space, our proposed methods only have a few key hyper-parameters that can significantly affect its performance. We investigate how will the number of layers $L$ and embedding size $D$ affect the recommendation performance of AutoMLP.

\begin{figure}[t]
     \centering
    {\subfigure{\includegraphics[width=0.49\linewidth]{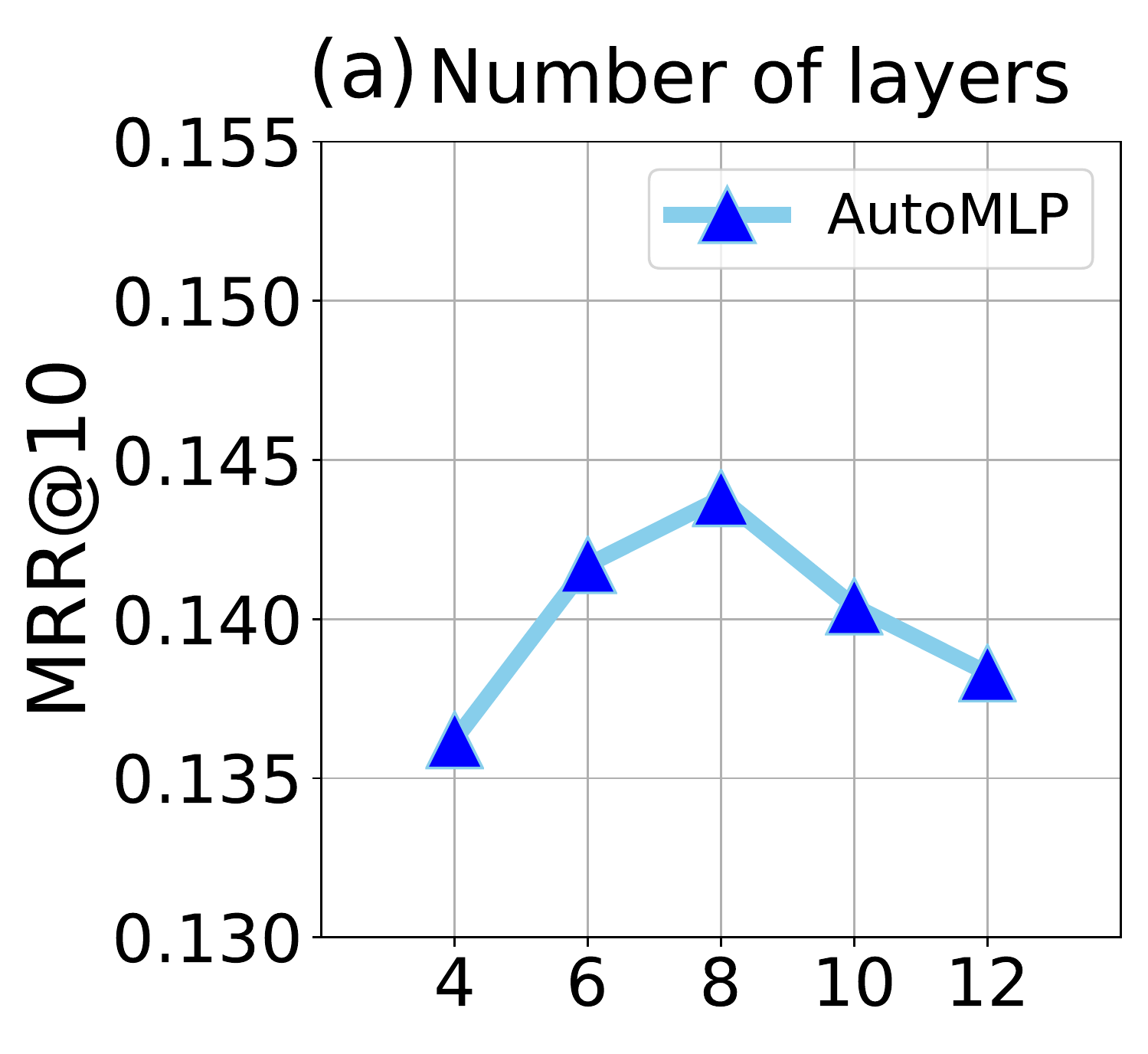}}}
    {\subfigure{\includegraphics[width=0.49\linewidth]{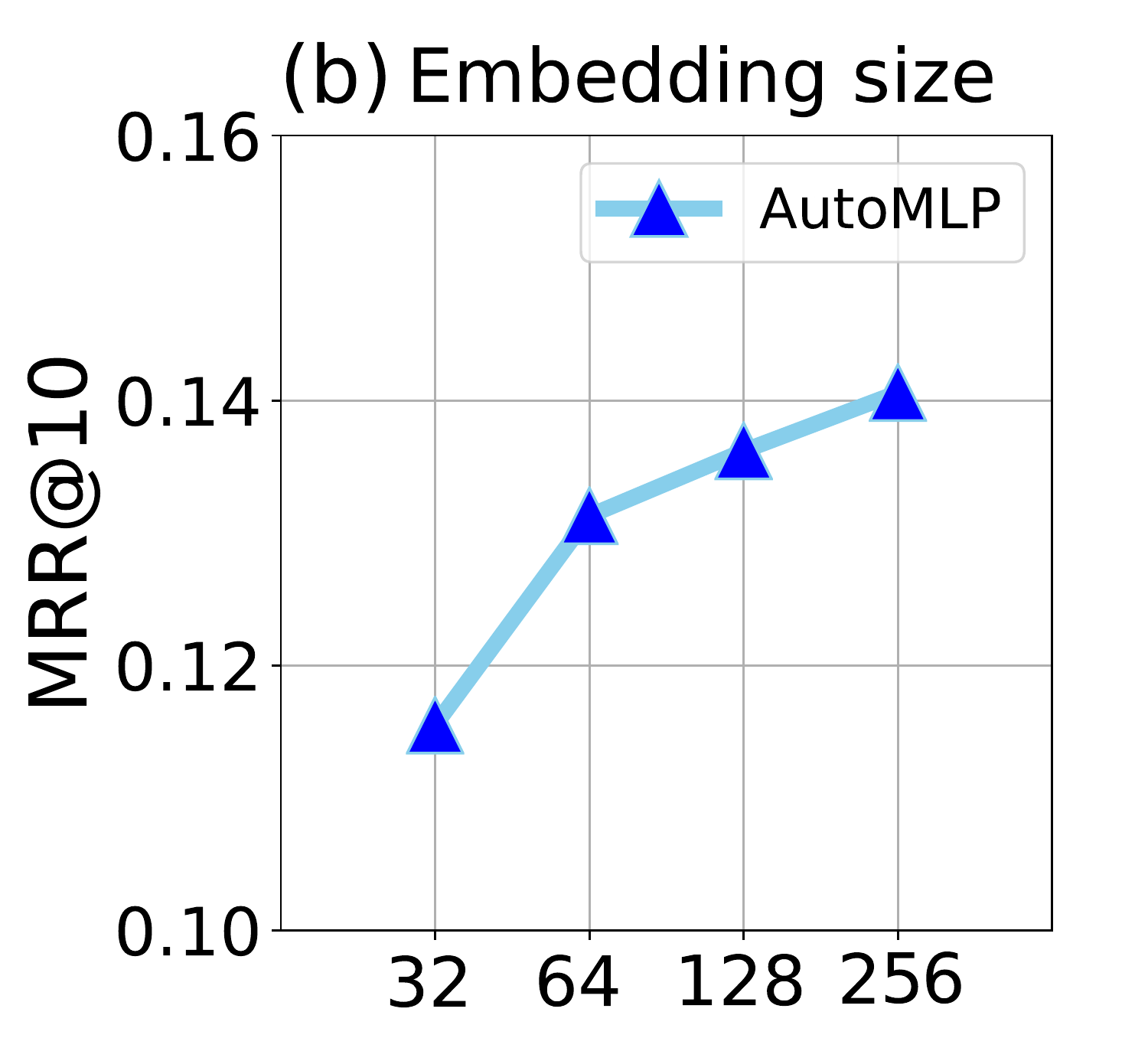}}}
    \vspace{-4.9mm}
    \caption{Influence of hyper-parameters on Beauty dataset}
    \label{fig:hyper}
    \vspace{-6mm}
\end{figure}

As shown in Figure \ref{fig:hyper} (a), we can observe that the optimal number of layers is 8. In other words, fewer layers (e.g., $L=4$) can degenerate model representation ability, while more layers (e.g., $L=12$) may lead to the overfitting issue.
From Figure \ref{fig:hyper} (b), it can be summarized that small embedding sizes will downgrade the model performance; with the increase of embedding size, AutoMLP possesses representation capacity to better learn the user-item interaction representations.

\subsection{Ablation Study}

In previous subsections, we have illustrated the effectiveness and efficiency of AutoMLP. An intuitive question behind such effectiveness is how we assign credits to each module of our proposed model. To validate the importance of each part of our model, we devise simplified alternatives architecture, each with one key component removed. Performances are reported in Table \ref{table:ablation}. Where w/o Sequence Mixer is removing Sequence Mixer from AutoMLP, and w/o Channel Mixer is to remove Channel Mixer from AutoMLP.

As shown in Table \ref{table:ablation}, we can observe that without Sequence-Mixer, the performance of AutoMLP drops most significantly. As we introduced in the Framework section, Sequence-Mixer aims to capture the sequential correlation of input sequence, without it the model will simply make predictions assuming input sequences are independent. The significant decrease indicates the core importance of sequential information in our recommendation task, and Sequence-Mixer can effectively capture such information.

When we only remove Channel-Mixer, we can also observe a drop in performance, but less significantly compared with removing Sequence-Mixer. The performance drop suggests that exchanging cross-channel information after each layer can also improve the quality of prediction by better learning the joint correlations of different implicit attributes of items.

\section{Related work}
\label{sec:relatedwork}
In this section, we briefly summarize several lines of works related to us, consisting of sequential recommendation and MLP-mixer.

\subsection{Sequential Recommendation}
Sequential recommendation plays an essential part in recommender systems. It aims at depicting users' successive preferences based on sequential interactions between users and items. 
RNN enjoys a natural strength in modeling sequential dependencies and takes the major part of the deep learning-based sequential recommendation. 
GRU4Rec \cite{hidasi2015session} employs the Gated Recurrent Unit (GRU) in the session-based recommendation. 
It's improved version \cite{hidasi2018recurrent} further boosts the performance with a ranking loss function as well as an improved sampling strategy.
However, RNN suffers from a high training cost, especially when modeling long-term sequences. 
Furthermore, both RNN-based models can not capture long-term dependencies well due to the vanishing gradient problem \cite{wang2019sequential}. 

Since the successful application in \cite{bahdanau2014neural}, attention has been widely applied to the sequential recommendation. 
NARM \cite{li2017neural} captures sequential user-item interactions as well as the user's main purpose in the current session through an encoder-decoder framework. 
Yu \etal \cite{yu2019multi} propose to maintain user preference through jointly learning the individual- and union-level item interactions. 

Lately, thanks to the advances in related areas \cite{vaswani2017attention}, the transformer has achieved better performance than RNN structure.
To mitigate the deficiency of RNN-based models, SASRec \cite{kang2018self} proposes to use an attention mechanism to model long-term sequential dependency. 
Bert4Rec \cite{sun2019bert4rec} employs deep bidirectional self-attention to learn user preferences.  
FDSA \cite{zhang2019feature} models transition patterns by integrating heterogeneous features with different weights. 
Besides, transformer-based models do not account for the order of input sequence without a heuristic process\cite{ke2020rethinking}.

\begin{table}[t]
\centering
\caption{Ablation study comparison on Beauty dataset}
\scalebox{1.1}{
\begin{tabular}{cccc} 
\hline
 Model & MRR@10 & NDCG@10 & HR@10  \\ 
\hline
w/o Sequence Mixer & 0.1132 & 0.1315 & 0.2169 \\
w/o Channel Mixer & 0.1408 & 0.1720 & 0.2700\\
\textbf{AutoMLP} & \textbf{0.1438} & \textbf{0.1754} & \textbf{0.2779}  \\
\hline
\end{tabular}}
\label{table:ablation}
\vspace{-2mm}
\end{table}

\subsection{MLP-mixer}

Recent advances in MLP architectures ~\cite{tolstikhin2021mlp,liu2021pay,touvron2022resmlp,tatsunami2021raftmlp} show that with simple alternations in design, MLP can serve as a strong alternative to Transformer-based models and attain competitive performances against state-of-the-art methods while maintaining both smaller time and space complexity. Among them, MLP-Mixer ~\cite{tolstikhin2021mlp} is usually considered the most representative work, by using separate MLP blocks to learn the per-location correlations and cross-location correlations from images, MLP-Mixer shows great scalability on large datasets and comparable performance against Transformer-based and Convolution Neural Network (CNN) based methods. 

We note that our proposed method is not the first endeavor to try to apply MLP-only architecture in a recommender system. MOI-Mixer~\cite{lee2021moi} first investigates the possibility of using MLP architecture as a substitute for Transformer-based methods. Subsequently, MLP4Rec~\cite{li2022mlp4rec}  proposed a tri-directional information fusion scheme, to coherently capture higher-order interactions across different attribute levels under a feature-rich scenario. 

\section{Conclusion}

In this paper, we proposed a long-term short-term sequential recommender system named AutoMLP. By only leveraging MLP architectures, AutoMLP shows competitive performances against state-of-the-art methods on both open-sourced benchmark datasets and real-world dataset from industrial application. Moreover, we devised an automated short-term interest length search algorithm that can efficiently learn an optimal short-term interest length, together with the linear complexity of MLP architectures, our method shows better efficiency and more promising improvement space compared with the existing methods. 

\clearpage

\section*{Acknowledgement}

This research was partially supported by APRC - CityU New Research Initiatives (No.9610565, Start-up Grant for New Faculty of City University of Hong Kong), SIRG - CityU Strategic Interdisciplinary Research Grant (No.7020046, No.7020074), HKIDS Early Career Research Grant (No.9360163), Huawei Innovation Research Program and Ant Group (CCF-Ant Research Fund).


\bibliographystyle{ACM-Reference-Format}
\bibliography{8_References}

\end{document}